# Simulink based VoIP Analysis

Hardeep Singh Dalhio, Jasvir Singh, M. Mian

**Abstract**— Voice communication over internet not be possible without a reliable data network, this was first available when distributed network topologies were used in conjunction with data packets. Early network used single centre node network in which a single workstation (Server) is responsible for the communication. This posed problems as if there was a fault with the centre node, (workstation) nothing would work. This problem was solved by the distributed system in which reliability increases by spreading the load between many nodes. The idea of packet switching & distributed network were combined, this combination were increased reliability, speed & responsible for voice communication over internet. Voice-over-IP (VoIP) These data packets travel through a packet-switched network such as the Internet and arrive at their destination where they are decompressed using a compatible Codec (audio coder/decoder) and converted back to analogue audio. This paper deals with the Simulink architecture for VoIP network.

━━━━━━━━━━━━━━ ◆ ━━━━━━━━━━━━━━

## 1 INTRODUCTION

The aim of voice transmission over IP is to find the technical solution of the problems which affects the QoS of VoIP network like delay, jitter, delay variation, packet loss, speech compression and offer a QoS under most network conditions. So our overall objective is to reach the high quality level of service provided by VoIP[1]. In a VOIP system there are three main components, speech collection / playing, CODEC (coder/decoder). Generally speaking, at the transmitter side, the speech signal is collected and encoded before transmitted to IP networks. At the receiver side, we do the reverse processes. The received data stream is decoded and recovered into speech signal and played back. The main function of speech collection is performed by software (C++) & hardware (sound card). [2]

## 2 REQUIREMENT

In the VoIP network there must be functionality similar to the one of the PSTN Network. To connect the VoIP network to the PSTN there is the need of a hardware called Gateway. Other terminals which includes IP phones and software running applications on computer. The LAN infrastructure, consisting of wires and switches, a gatekeeper which manages the terminals in a zone, providing address resolution, registration, authentication and regulating admission to the VoIP system; optional call manager software applications running on the gatekeeper or on a standalone platform; a router, which connects LAN segments together and to WAN circuits; gateways to perform format conversions between the VoIP world and the PSTN; and optional connections to the PSTN, the Internet and to managed IP WAN services or IP VPN services. A firewall is needed when connecting to the Internet as shown in fig.1 [3-4]

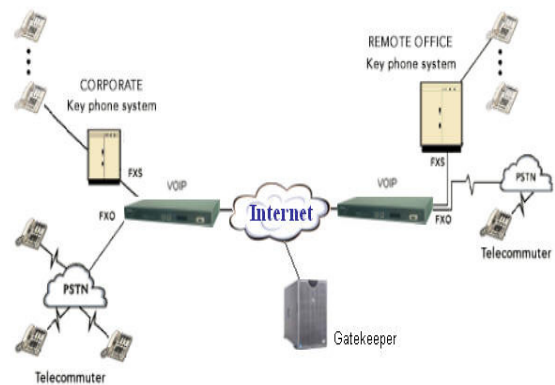

Fig .1 VoIP Network [4]

- *Hardeep Singh is Research Scholar Department of Physics, Guru Nanak Dev University, Amritsar.*
- *Dr. Jasvir Singh is Professor in Dept of Electronics Technology, Guru Nanak Dev University, Amritsar.*
- *Prof. M. Mian is Professor in Dept of Physics, Guru Nanak Dev University, Amritsar.*

## 3 A SIMULINK INTERFACE FOR VOIP ANALYSIS

Here Fig.2 shows the Simulink interface used for sending and receiving data transmission over IP network. Using this interface we have tried to built an VoIP network based on Simulink. [5-6]

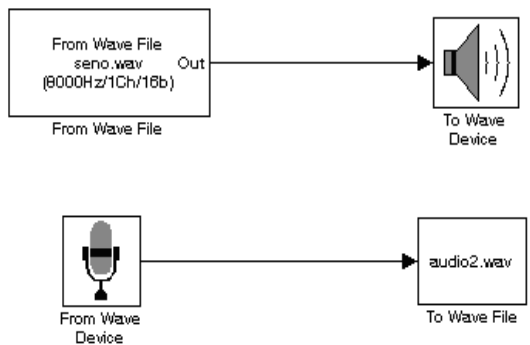

Fig 2 VoIP Simulink Interface [5]

## 4 VOIP SIMULINK MODEL

Here we built a voip environment for real time processing, Fig.3 shows the simulink model for VoIP using MATLAB version 7.3. Model describes the transmitter portion of a VoIP system. In this model we are giving speech signal from wave file speech_dft 8khz, here we are using G.711 codec, the selection of codec depends upon system needs. In this model we can see the variation in input signal through spectrum and time scope. Using this model we can verify various parameters affecting the QoS of voice over IP. This block consists of wave file as the input to the codec through gain. The generated signal is further given to UDP for onward transmission. The spectrum analysis is done using time scope, power density, periodogram, packet spectrogram we analyze the signal generation and its parameters.

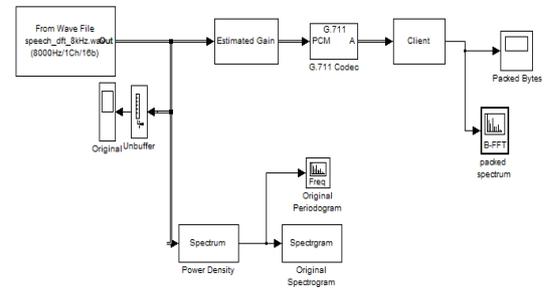

Fig 3 VoIP Transmission over UDP

Similarly, in this part we have receive the VoIP signal at UDP receiver as shown below in Fig.4

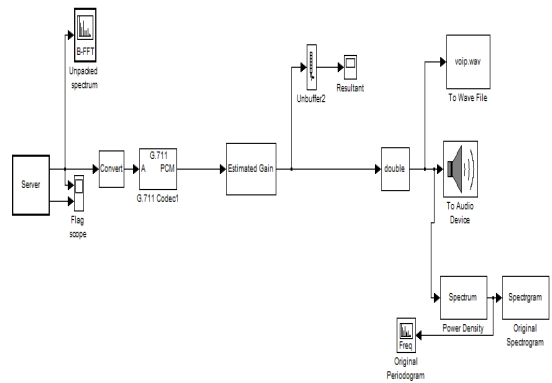

Fig. 4. VoIP Reception from UDP

## 5 RESULTS AND DISCUSSION

The VoIP wave is generated using VoIP Simulink interface, the frequency spectrum of the original voip signal is shown in Fig 5. The spectrogram of this signal is shown in Fig. 6 the periodogram of the original signal is shown in Fig.7 and before sending the signal for transmission over UDP it is packetized. Fig. 8 shows the packed spectrum of the VoIP signal.

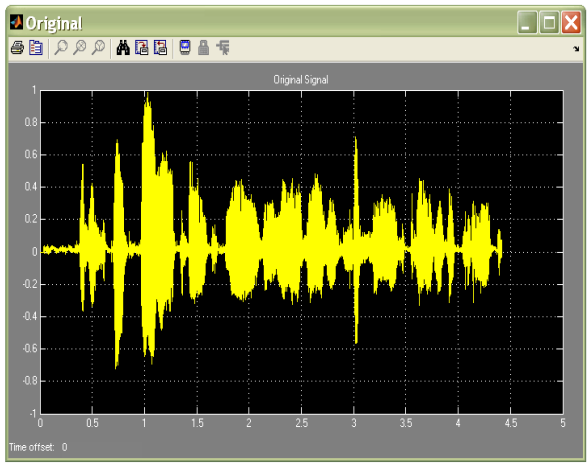

Fig .5 Original Signal

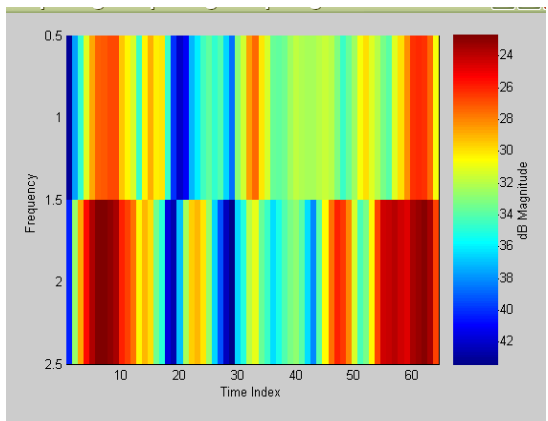

Fig. 6 Spectrogram of Original Signal

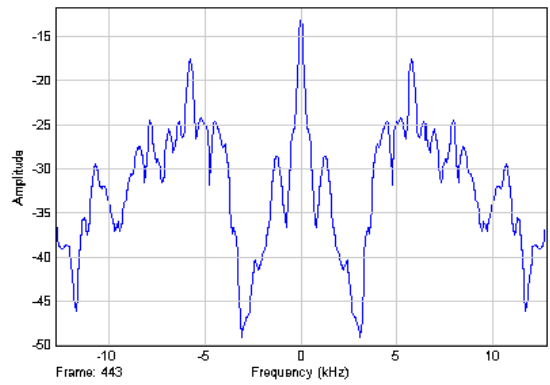

Fig. 7 Original Periodogram

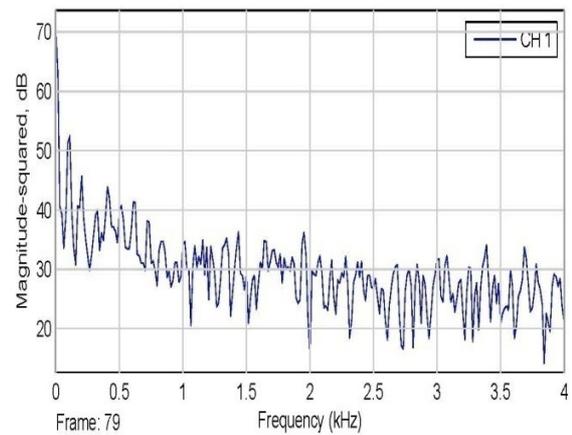

Fig. 8 Packed Spectrum of the VOIP Signal

The output of VoIP receiver model is observed at UDP receiver end and Fig. 9, Fig. 10, Fig.11 and Fig.12 shows the unpacked spectrum, resultant signal in frequency domain, period gram and spectrogram results obtained at the receiver part of the VoIP network.

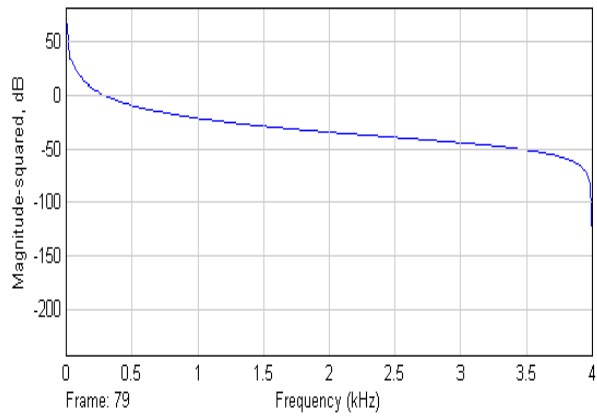

Fig.9 Unpacked Spectrum at UDP Reciever end

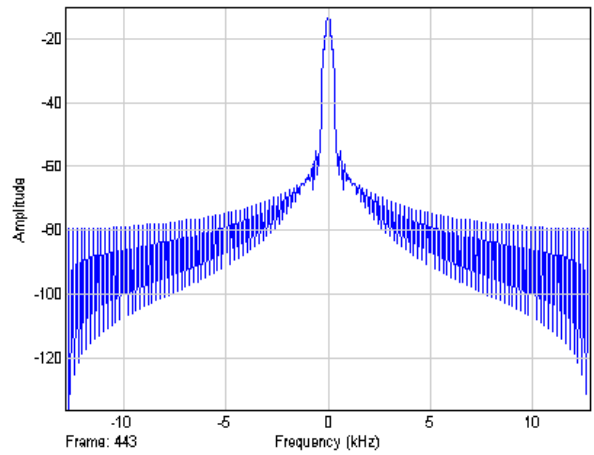

Fig. 11 Original Periodogram

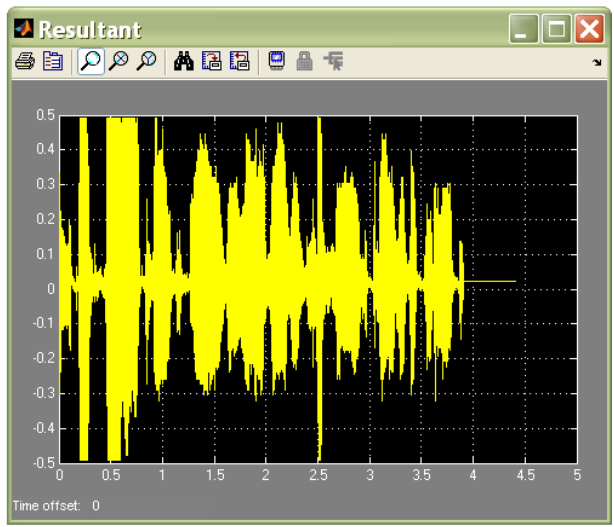

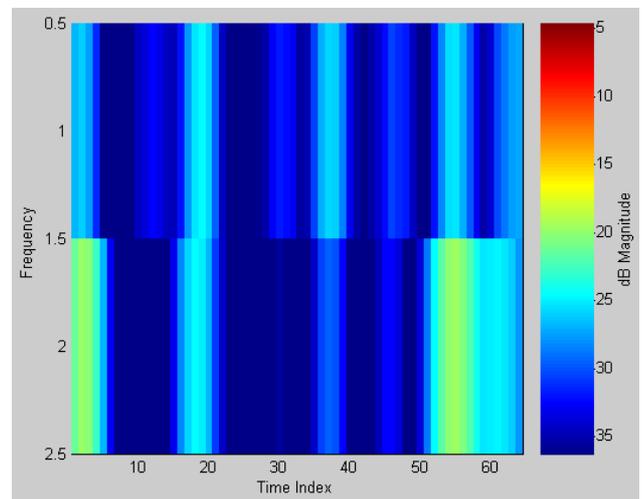

Fig.12 Original Spectrogram of VOIP Network

Fig.10 Resultant at Reciever end of VoIP network

## 6 CONCLUSION

The present paper has highlighted the role of VoIP in the present scenario the Simulink models have been developed for VoIP environment. The frequency and spectrum analysis for different wave forms for both the transmitter and receiver has been presented in VoIP environment. Study has impact in the design and analysis Of Adaptive VoIP system to improve the performance the work is in progress for incorporating more variables in the existing system for future VoIP applications.